\documentclass[a4paper,11pt]{article}
\usepackage{pos}
\title{Modeling the non-flaring VHE emission from M87 as detected by the HAWC gamma ray observatory}
 \ShortTitle{Modeling the non-flaring VHE emission from M87}

\author*[a]{Fernando Ureña-Mena}
\author[a]{Alberto Carramiñana}

\author[b]{Anna Lia Longinotti}
\author[a]{Daniel Rosa-González}

\affiliation[a]{Instituto Nacional de Astrofísica, Óptica y Electrónica (INAOE)\\
  Luis Enrique Erro 1, Tonantzintla, Puebla, Mexico}

\affiliation[b]{Instituto de Astronomía, Universidad Nacional Autónoma de México (IA-UNAM),\\ Ciudad de México, Mexico\\
}

\forColl{HAWC} 

\emailAdd{furena@inaoep.mx}

\abstract{M87 is a giant radio galaxy located in the Virgo Cluster, known to be a very high energy (VHE) gamma-ray source. As radio galaxies are considered the misaligned low-redshift counterparts of blazars, they are excellent laboratories for testing AGN emission models. M87 has been detected and monitored by Fermi-LAT and several atmospheric Cherenkov telescopes. Recently, the HAWC Collaboration presented  marginal evidence of 4.5 year TeV gamma-ray emission from this object. The HAWC data has the potential to constrain the average VHE emission of sources of complex behavior, like M87, for which the physical origin of the VHE gamma-ray emission is still uncertain. We fitted a lepto-hadronic scenario to the broadband spectral energy distribution of M87 to model its non-flaring VHE emission using HAWC data. }

\FullConference{37$^{\rm{th}}$ International Cosmic Ray Conference (ICRC 2021)\\
		July 12th -- 23rd, 2021\\
		Online -- Berlin, Germany}

\begin{document}
\maketitle

\section{Introduction}

M87 is a supergiant elliptical galaxy with an active  nucleus (AGN). It is classified as a radio galaxy (RDG) of type Fanaroff-Riley I (FR-I) and is located in the Virgo Cluster at a distance of 16.7 $\pm$ 0.2 Mpc \cite{mei2007}. Its prominent jet was first discovered in 1918 \cite{curtis1918} and has been studied in multiple wavelengths and scales \cite{blandford2019}.  The Event Horizon Telescope (EHT) observed the supermassive black hole (SMBH)  M87*, which engines the AGN of M87. producing an image of its shadow \cite{ehtI2019}. This result was used to constrain the black hole mass  \cite{ehtVI2019}, spin and recently, the magnetic field structure near the event horizon \cite{ehtVIII2021}. \\

M87 is a well-established MeV, GeV and TeV gamma-ray source \cite{abdo2009,magic2020}. It was the first RDG detected in the TeV gamma-ray range and some TeV flares have been observed \cite{aharonian2006,albert2008,abramowski2012}. In non-flaring state, this source has also shown complicated gamma-ray spectral and flux variability \cite{benkhali2019}. The zone where this emission is produced is not well determined, being the inner jet or core its most likely origin \cite{abramowski2012,benkhali2019}. Other candidates are the jet feature HST-I and the SMBH vicinity \cite{abramowski2012,benkhali2019}. The physical mechanism that produces this emission is not known either. It is commonly accepted that an one-zone synchrotron self Compton (SSC) scenario is not enough to explain this emission i.e, \citep{fraija2016,georganopoulos2005}. This is supported by the evidence of a spectral turnover at energies of $\sim 10$ GeV, which could be produced by the transition between  two different emission processes \cite{benkhali2019}. One of the proposed alternatives to explain this emission are the lepto-hadronic scenarios, which explain the  spectral energy distribution (SED) combining leptonic models with photo-hadronic interactions \cite{fraija2016,sahu2015}.\\

The High Energy Water Cherenkov (HAWC) gamma-ray observatory marginally detected this source at $E>0.5$ TeV \cite{albert2021}. This long-term 1523 days observation represents a good constraint of the average TeV emission of M87. In this work we fit a lepto-hadronic model to a SED built to  include the HAWC observations for the first time. \\

\section{Data}

An average SED of M87 was constructed using historical archive data \cite{morabito1986,morabito1988,junor1995,lee2008,lonsdale1998,doeleman2012,biretta1991,perlman2001,sparks1996,marshall2002,wong2017,abdo2009}. We also included \textit{Fermi}-LAT observations from the 4FGL catalog \cite{abdollahi2020}. As it was mentioned above, HAWC data from \cite{albert2021} were used to consider the VHE emission. These observations were already corrected by an Extra-galactic background light (EBL) absorption model \cite{dominguez2011}. Therefore, we do not consider this effect during the fitting process.  

\section{Model and Methodology}
The model that we use postulates an electron population contained in spherical region in the inner jet \cite{finke2008}. The spectral electron distribution (as a function  of $\gamma^\prime$, the  electron comoving Lorentz factor) is a broken  power law given by Equation \ref{eq:Ne} :

\begin{equation}
    N_e(\gamma^\prime)\propto \begin{cases} \gamma^{\prime^{-p_1}} \text{for } \gamma^\prime<\gamma_c^\prime \\  
     \gamma^{\prime^{-p_2}} \text{for } \gamma^\prime>\gamma_c^\prime
    
    \end{cases},
    \label{eq:Ne}
\end{equation}

where $p_1$,$p_2$ correspond to the power law indices, and $\gamma^\prime_c$ to the break Lorentz factor, which is one of the fitting  parameters of the model. Other two model parameters are B, the magnetic field intensity and $D$, the Doppler factor of the emission zone. The one-zone SSC scenario explains the SED range from radio to X-rays as synchrotron emission produced by electrons moving in  the magnetic field. A second energy component, from X-rays to gamma rays, is produced when  electrons Compton scatter synchrotron photons \cite{finke2008}. \\

The proton population in the emission zone is assumed to have a single power law spectral distribution (Equation \ref{eq:Np}):

\begin{equation}
\label{eq:Np}
N_p(\gamma_p^\prime) \propto \gamma_p^{\prime-\alpha},
\end{equation}

where $\alpha_p$ is the proton spectral index, which is a fitting parameter of the model. This scenario postulates that gamma-ray emission is produced in particle cascades generated by interaction between SSC  photons and accelerated protons \cite{sahu2019}. The other fitting parameter of this model is a normalization constant $A_\gamma$ \cite{sahu2019}. \\

The methodology consisted of developing a Python code to fit the emission model to the SED. First, the one-zone SSC model \cite{finke2008} was fitted to the data between radio and MeV-GeV gamma rays. Then, the photo-hadronic \cite{sahu2019} component was added to fit the VHE emission.  The best fit values of the model parameters were obtained with chi-square minimization and errors were estimated with Monte Carlo simulations. \\

\section{Results and Discussion}

\begin{figure}
    \centering
    \includegraphics[width=\textwidth]{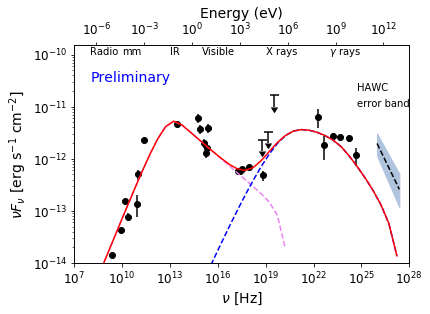}
    \caption{Spectral Energy Distribution (SED) of the RDG M87 with the SSC model fit. Archive data points are shown black. The  violet curve corresponds to the synchrotron component and  the blue curve to the inverse Compton one.  The red curve corresponds to the total emission. The HAWC 1$\sigma$ error band is shown in light blue.  }
    \label{fig:SSC}
\end{figure}
    
\begin{figure}
    \centering 
    \includegraphics[width=\textwidth]{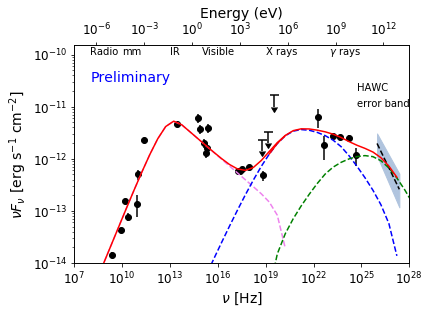}
    \caption{Spectral Energy Distribution (SED) of the RDG M87, including HAWC data, with the lepto-hadronic model fit.  Archive data points are shown black.  The  violet curve corresponds to the synchrotron component,  the blue curve to the inverse Compton component and the green curve to photo-hadronic one. The red curve corresponds to the total emission. The HAWC 1$\sigma$ error band is shown in light blue. }
    \label{fig:ph}
\end{figure}
    
\begin{table}[]
    \centering
    \begin{tabular}{c c c} \hline
           Parameter& & Value \\ \hline
          Magnetic Field intensity (mG) & $B$&$ 46 \pm 3$  \\
  Doppler Factor & $D$ & $ 4.3\pm 0.2$\\ \hline
  \textbf{Electron spectral parameters} \\ \hline
  Broken  PL index & $p_1$ &$1.52 ^{+0.02}_{-0.01} $\\
 Broken PL index & $p_2$ & $ 3.53 \pm 0.02 $\\
    Break Lorentz factor ($\times 10^3$) & $\gamma_c^\prime$ &$ 3.80 ^{+0.07}_{-0.05} $   \\ \hline
   \textbf{Photo-hadronic Component} \\ \hline 
    Proton spectral index & $\alpha$ & $3.0 \pm 0.2 $ \\ 
    Normalization & $\log(A_\gamma)$ & $-0.5\pm 0.2  $ \\ \hline
    $\chi^2_\nu$(d.o.f) & & 25.8 (22) \\ \hline
    \end{tabular}
    \caption{Best fit values of the fitting parameters}
    \label{tab:SSC}
\end{table}

The best fit model for the SSC scenario is shown in Figure \ref{fig:SSC}. At first glance, it seems that the VHE spectral region shows a spectral hardening which cannot be explained with the SSC components. This agrees with the spectral turnover  reported by \cite{benkhali2019},  which would explain that the VHE SSC flux prediction is well below the HAWC observed spectrum.\\

The best fit model for the entire lepto-hadronic scenario is shown in Figure \ref{fig:ph}. As can be seen, this component is necessary to fit both  GeV and TeV  gamma-ray observations. A slight spectral hardening is seen at $E\sim10$ GeV, which agrees with the results obtained in \cite{benkhali2019}. The fit results are shown in Table \ref{tab:SSC}. \\

A similar spectral hardening is also observed in the gamma-ray emitter RDG Centaurus A  \cite{sahakyan2013}.  The existence of an additional emission component in TeV emitter RDGs could be expected due to their low Doppler factors ($D \lesssim 10$)\cite{rieger2018}, compared to those of blazars whose gamma-ray emission is enhanced by their substantial  Doppler-boosting.  Moreover, the possible detection of neutrino emission from blazars suggests that similar mechanisms may be present in other types of AGNs  \cite{aartsen2018}. \\

The goal of this  work is explaining the VHE emission of M87, but specific features in the rest of the bands need a more detailed analysis \cite{algaba2021}.  Multi-zone models are probably needed, as well as some specific data sets. In case of the mm bands, the recent EHT results could be very helpful.\\

HAWC data correspond to the first long-term continuous TeV observation of this source. Previous air Cherenkov campaigns constrained very well the average TeV emission, however they could be affected by short-term spectral and flux variations. For example, in the MAGIC two-year campaign reported in \cite{acciari2020} no evidence of the gamma-ray spectral turnover was found. However, this campaign was coincident with a high activity period reported by \cite{benkhali2019}, where the spectral turnover seems to have temporarily disappeared. A more detailed analysis would be needed to know if these short term variations could be explained with the photo-hadronic scenario.\\ 

HAWC is still continuously taken data, which can be used in a near future to improve the results of this analysis, as well as to test other physical scenarios. 

\subsection{Summary and conclusions}

M87 is a giant RDG that emits in gamma rays up to TeV bands. The physical mechanism that produces the VHE emission has yet to be determined. The lepto-hadronic scenario explains this emission with photo-hadronic interactions, whereas the rest of the broadband  SED is attributed to a leptonic mechanism. The HAWC Collaboration recently reported a marginal detection of this source that can be used to constrain its average VHE emission, which can be used to test possible physical scenarios. In this work we fit a lepto-hadronic model to a  SED which includes the results from HAWC.  We obtained best  values for the fitting parameters, which are the mean magnetic field intensity ($B=46\pm3$ mG), the Doppler factor ($D=4.3\pm 0.2$), the electron spectral parameters ($p_1=1.52^{+0.02}_{-0.01},p_2=3.53\pm0.02,\gamma_c^\prime=3.80^{+0.07}_{-0.05} \times 10^{3}$) and the photo-hadronic parameters ($\alpha=3.0\pm0.2$, $\log(A_\gamma)=-0.5\pm0.2$). We concluded that this scenario could explain the M87 VHE emission, including some spectral features like a possible turnover at $\sim 10$ GeV. 

 \acknowledgments
 We acknowledge the support from: the US National Science Foundation (NSF); the US Department of Energy Office of High-Energy Physics; the Laboratory Directed Research and Development (LDRD) program of Los Alamos National Laboratory; Consejo Nacional de Ciencia y Tecnolog\'ia (CONACyT), M\'exico, grants 271051, 232656, 260378, 179588, 254964, 258865, 243290, 132197, A1-S-46288, A1-S-22784, c\'atedras 873, 1563, 341, 323, Red HAWC, M\'exico; DGAPA-UNAM grants IG101320, IN111716-3, IN111419, IA102019, IN110621, IN110521; VIEP-BUAP; PIFI 2012, 2013, PROFOCIE 2014, 2015; the University of Wisconsin Alumni Research Foundation; the Institute of Geophysics, Planetary Physics, and Signatures at Los Alamos National Laboratory; Polish Science Centre grant, DEC-2017/27/B/ST9/02272; Coordinaci\'on de la Investigaci\'on Cient\'ifica de la Universidad Michoacana; Royal Society - Newton Advanced Fellowship 180385; Generalitat Valenciana, grant CIDEGENT/2018/034; Chulalongkorn University’s CUniverse (CUAASC) grant; Coordinaci\'on General Acad\'emica e Innovaci\'on (CGAI-UdeG), PRODEP-SEP UDG-CA-499; Institute of Cosmic Ray Research (ICRR), University of Tokyo, H.F. acknowledges support by NASA under award number 80GSFC21M0002. We also acknowledge the significant contributions over many years of Stefan Westerhoff, Gaurang Yodh and Arnulfo Zepeda Dominguez, all deceased members of the HAWC collaboration. Thanks to Scott Delay, Luciano D\'iaz and Eduardo Murrieta for technical support.

%
%
%
\clearpage
\section*{Full Authors List: \Coll\ Collaboration}

\scriptsize
\noindent
A.U. Abeysekara$^{48}$,
A. Albert$^{21}$,
R. Alfaro$^{14}$,
C. Alvarez$^{41}$,
J.D. \'Alvarez$^{40}$,
J.R. Angeles Camacho$^{14}$,
J.C. Arteaga-Vel\'azquez$^{40}$,
K. P. Arunbabu$^{17}$,
D. Avila Rojas$^{14}$,
H.A. Ayala Solares$^{28}$
% ...

\noindent
% ...
$^{14}$Instituto de F\'isica, Universidad Nacional Aut\'onoma de M\'exico, Ciudad de M\'exico, M\'exico,
% ...
$^{17}$Instituto de Geof\'isica, Universidad Nacional Aut\'onoma de M\'exico, Ciudad de M\'exico, M\'exico,
% ...
$^{21}$Physics Division, Los Alamos National Laboratory, Los Alamos, NM, USA,
% ...
$^{28}$Department of Physics, Pennsylvania State University, University Park, PA, USA,
% ...
$^{40}$Instituto de F\'isica y Matem\'aticas, Universidad Michoacana de San Nicol\'as de Hidalgo, Morelia, Michoac\'an, M\'exico,
$^{41}$FCFM-MCTP, Universidad Aut\'onoma de Chiapas, Tuxtla Guti\'errez, Chiapas, M\'exico,
% ...
$^{48}$Department of Physics and Astronomy, University of Utah, Salt Lake City, UT, USA,
% ...

\end{document}